# Electronic states of LaCoO$_3$: Co K-edge and La L-edge X-ray absorption studies

S. K. Pandey[*], Ashwani Kumar[a], S. Khalid[b], and A. V. Pimpale[§]

UGC-DAE Consortium for Scientific Research, University Campus, Khandwa Road, Indore 452 017, India

[a]School of Physics, Devi Ahilya University, Khandwa Road, Indore 452 017, India

[b]National Synchrotron Light Source, Brookhaven National Laboratory, Upton, New York – 11973

**Abstract**

Room temperature Co K- and La L-edges X-ray absorption studies have been carried out on LaCoO$_3$. Experimental near edge structures have been analysed by theoretical LDA+U density of states (DOS) and multiple scattering (MS) calculations. Use of both MS and DOS calculations yield additional information about hybridization of the states of central atom with neighbouring atoms responsible for producing the near edge structures. Absorption processes at Co K, La L$_1$, L$_2$, and L$_3$-edges have been attributed to electronic transitions from Co 1s$\rightarrow$ Co 4p, La 2s$\rightarrow$ La 6p, La 2p$_{1/2}$$\rightarrow$ La 5d, and La 2p$_{3/2}$$\rightarrow$ La 5d, respectively. All the pre-edge and post-edge features including the shape of the main absorption edge have been generated by taking the convolution of the calculated DOS, indicating that single particle approximation is sufficient to express all experimentally observed major structures. Two pre-edge structures observed in Co K-edge spectrum are attributed to Co 1s$\rightarrow$ $e_g^\uparrow$ and $e_g^\downarrow$ quadrupole transitions in contrast to earlier identification of the same to Co 1s$\rightarrow$ t$_{2g}$ and e$_g$ transitions. The influence of La 6p states on the Co 4p states is such that the inclusion of La atoms in the MS calculations is necessary to generate post-edge structures in Co K-edge spectrum. The importance of the hybridization of O 2p state with La 6p and 5d in the L-edge absorption processes has also been discussed. 10% contribution of quadrupole channel has been estimated in La L-edges.

**Introduction**

Cobaltates having general formula ACoO$_3$ (A-rare earth element) have been of great interest among the researchers for last fifty years as they show temperature induced insulator to metal and non-magnetic to paramagnetic transitions [1]. Among the cobaltates LaCoO$_3$ is a relatively simple system as its magnetic property is mainly decided by Co$^{3+}$ ion and the La$^{3+}$ ion with completely filled orbitals play little role. Thus, to understand the basic physical mechanism responsible for temperature-induced transitions, LaCoO$_3$ is a good candidate. Temperature dependent magnetic susceptibility data for this compound show a dip around 35 K and two peaks around 100 K and 550 K, respectively [2]. Earlier it was believed that such temperature-induced transitions occur due to spin state transitions of Co$^{3+}$ from low spin (LS) $t_{2g}^6 e_g^0$ state to high spin (HS) $t_{2g}^4 e_g^2$ state [3]. Later, Potze et al. introduced the concept of intermediate spin (IS) $t_{2g}^5 e_g^1$ state in these compounds [4] and showed that such states play important role in these transitions. Currently, it is generally believed that the first magnetic transition around 100 K is due to spin state transition of Co$^{3+}$ ion from LS to IS and the second one around 550 K due to IS to HS transition.

Electronic states of LaCoO$_3$ are decided by the competition between the Hunds coupling energy and crystal field energy. Since the difference in the values of these two physical quantities is typically ~ k$_B$T, these states are rather sensitive to temperature. Thornton et al. [5] have carried out temperature dependent Co K-edge X-ray absorption spectroscopy (XAS) studies on LaCoO$_3$. They have seen two pre-edge structures and attributed the same to Co 1s→3d quadrupole transition involving the split t$_{2g}$ and e$_g$ orbitals. Abbate et al. [6] have carried out temperature dependent Co 2p and O 1s XAS studies on this compound. Their studies revealed that at temperature below 300 K LaCoO$_3$ was in highly covalent LS state and at temperature above 550 K it was in mixed LS and HS spin states. Saitoh et al. [7] have also carried out O 1s XAS studies including Co 2p core-level photoemission and the valence band photoemission on this compound. They showed that the ground state of this compound was LS state with heavily mixed $d^6$ and $d^6 \underline{L}$ configurations. They have attributed the transition around 100 K to LS to IS transition. Wu et al. [8] have also studied Co and O K-edge XAS to see the effect of core-hole potential and structural distortion on XAS spectra. Recently, Medarde et al. [9] have carried out systematic temperature dependent Co K$_{\alpha 1}$ XAS studies on LaCoO$_3$ to investigate the low temperature spin state transition. Their data show two pre-edge structures between 7708 and 7712 eV. Based on earlier reports [10] they have attributed first and second peaks to the transitions Co 1s→t$_{2g}$ and 1s→e$_g$ levels, respectively, similar to ref. 5. Their studies revealed the possibility of magnetic ground state below 30 K in contrast to generally believed non-magnetic ground state.

XAS is an important experimental tool to study the electronic structure of a compound. It gives information about unoccupied states. Although some work on O and Co K-edges exist in the literature, very little work is seen on La L-edges for LaCoO$_3$. Some work on La L-edges XANES studies have been carried out for intermetallic system LaGe$_2$ [11]. For cobaltate systems rare earth L-edge studies are not available in literature to the best of our knowledge. K-edge mainly gives information about p-symmetric unoccupied states. Thus it cannot be very useful in extracting the information about d and f symmetric states. To get the full picture of electronic states of LaCoO$_3$ it is imperative to carry out La L-edge study as well.

Moreover, identification of pre-edge structures in this compound is based on available works on other similar compounds. It is well known that the theoretical DOS of cobaltates depends on initial electronic states used to calculate the electronic structure [12]. Thus the attribution of pre-edge structures to 1s→$t_{2g}$ and $e_g$ transitions may be rather crude for these compounds as the majority and minority spin occupancies are not taken into consideration. In this context, XAS studies of Co K- and La L-edges would yield complete information of the electronic structure of $LaCoO_3$.

We report here the room temperature Co K, La $L_1$, $L_2$ and $L_3$-edges XAS studies of $LaCoO_3$. To the best of our knowledge, the La L-edges in $LaCoO_3$ have been reported here for the first time. The experimental spectra are compared with the calculated ground state density of states (DOS) using LDA+U method. It is seen that all the features of the experimental spectra can be accounted for wthin this approximation. Two pre-edge structures are observed in the Co K-edge spectrum. Our studies indicate their origin to the quadrupole allowed Co 1s→$e_g^\uparrow$ and $e_g^\downarrow$ transitions, respectively. The multiple scattering (MS) analysis of post-edge features and their comparison with calculated DOS indicates the importance of hybridization of Co 4p state with La 6p in generating these features. However, in the case of La L-edges, it is the hybridization of La 6p and 3d states with O 2p states, which is important for understanding the post-edge structures. Moreover, there is about 10% contribution of quadrupole channel in the main-edge regions of the L-edges.

**Experimental**

A single-phase polycrystalline powder sample of $LaCoO_3$ was prepared by combustion method [13]. It was characterized by powder X-ray diffraction (XRD) technique. Lattice parameters obtained from Rietveld analysis match well with literature. Room temperature Co K- and La L-edge XAS experiments were done at beamline X-18 B at the National Synchrotron Light Source, Brookhaven National Laboratory. The storage ring was operated at 2.8 GeV, 300 mA. The beamline used a Si (111) channel cut monochromator. The horizontal acceptance angle of the beam at the monochromator was 1 mrad. The vertical slit size used in this experiment was 1 mm, corresponding to an energy resolution of about 0.8 eV at the Co K-edge. The average photon flux for this bandwidth was $10^{10}$ photons/sec. The monochromator was detuned by 35 % to reduce the higher harmonics. The incident ($I_0$) and the transmitted beams ($I_t$) were measured by sealed ion chambers, with the combination of gases for appropriate absorption. Standard Co foil was placed between the detectors $I_t$ and $I_{ref}$ for energy reference and to check the stability of the beamline and optical system. The sample powder sieved through a 400 mesh and again grinded with mortar and pestle was spread uniformly on a cellophane tape and a four-fold of this tape was used to minimize the pinhole and brick effects.

**Computational details**

The experimental absorption spectra were analysed using DOS and MS calculations. Spin polarized LDA+U electronic structure calculations were carried out using LMTART 6.61 [14]. For calculating charge density, full-potential LMTO method working in plane wave representation was used. In the calculation we have used the muffin-tin radii of 3.384, 1.951, and 1.661 a.u. for La, Co, and O sites, respectively. The charge density and effective potential were expanded in spherical harmonics up to *l* = 6 inside the sphere and in a Fourier series in the interstitial region. The initial basis set included 6s, 6p, 5d, and 4f valence and 5s semicore orbitals of La; 4s, 4p, and 3d valence and 3p semicore orbitals of Co and 2s and 2p valence orbitals of O. The exchange correlation functional of the density functional theory was taken after Vosko, Wilk, and Nussair and the generalized-gradient approximation scheme of Perdew et al. [15] was also invoked. In the LDA+U method three input parameters are required for 3d electrons. They are Slater integrals $F^0$, $F^2$, and $F^4$. These integrals are directly related with on-site Coulomb interaction (U) and exchange interaction (J) by relations $U = F^0$, $J = (F^2 + F^4) / 14$, and $F^4/F^2 \sim 0.625$. We have used U = 3.5 eV and J = 1.0 eV in our earlier calculation for $PrCoO_3$ as they yield results in conformity with experimental data [16]. We have adopted the same values for $LaCoO_3$ as changing Pr to La is not expected to make major changes in the Co-site. At room temperature $LaCoO_3$ is believed to be in IS state. Thus we started our calculation from IS state configuration, which corresponds to $t_{2g}^{3\uparrow}$, $e_g^{1\uparrow}$, and $t_{2g}^{2\downarrow}$. (6, 6, 6) divisions of the Brillouin zone along three directions for the tetrahedron integration were used to calculate the DOS. Self-consistency was achieved by demanding the convergence of the total energy to be smaller than $10^{-5}$ Ry/cell. The final self-consistent configuration remains in IS as expected. Full multiple scattering calculations were carried out using FEFF8.2 [17] in order to simulate XA spectra. It may be remarked that in LDA+U the potential is explicitly dependent on orbital degrees of freedom; whereas in FEEF8.2 which uses LDA, these orbital degrees of freedom are averaged out. Potentials for each atomic scatterer were calculated self-consistently for a cluster of atoms lying within a radius of ~ 5.5 Å. These potentials were then used to calculate the XA spectra for different cluster size up to a radius of ~ 5 Å. Both dipole and quadrupole absorption processes were considered in the calculations.

**Results and discussion**

### Co K-edge

In figure 1 experimental Co K-edge X-ray absorption (XA) spectrum of $LaCoO_3$ is shown by open circles. It is evident from the figure that the pre-edge region lies between 7707.5 and 7714.5 eV and the main edge between 7714.5 and 7726.5 eV. There are two pre-edge structures at 7710.3 and 7711.9 eV denoted by A and B, respectively; these are shown on expanded scale in the lower panel of the inset to this figure. In the main-edge region there are two features indicated by C and D and in the post-edge region there are three structures denoted by E, F, and G. Following usual convention we define the edge position as the first inflection point on the main absorption edge. The edge position of cobalt metal foil is taken at 7709 eV.

In the inset of figure 1, $t_{2g}$ and $e_g$ DOS are shown. It is clear from the figure that there is a finite DOS at the Fermi level in the majority spin channel thus giving the half-metallic ground state contrary to observed insulating behaviour. The IS configuration used in the calculations is known to yield a half-metallic state [12]. It is necessary to include orbital ordering due to Jahn-Teller active $Co^{3+}$ ion in IS state to create the required band gap in both the spin channels and make the system insulating. The inclusion of this effect shifts the Co $e_g$ and O 2p bands to somewhat higher energies, thus creating the necessary gap [12]. Further the self-consistent results indicate that there are about 6.8 electrons in the d-band. The extra 0.8 electrons are coming from the ligand due to hybridization between O 2p and Co 3d orbitals. These extra 0.8 electrons are in $e_g$ orbitals, thus the ratio of unoccupied $t_{2g}$ and $e_g$ DOS would be 1:2.2. The calculated ratio including the $t_{2g}$ DOS around 21 eV (not shown in figure 1) comes out to be 1:2.3 in conformity with expectation.

Under dipole approximation the K-edge absorption process occurs due to transition of electron from s-symmetric state to p-symmetric states. Our LDA+U electronic structure calculations indicate that there are finite DOS of Co 4p, O 2p, and La 6p states above the Fermi level in the appropriate energy range. Since O and La atoms are located at different sites, contribution from O 2p and La 6p states to Co K-edge absorption process would be negligible. Thus Co K-edge absorption can be identified to the electronic transition from Co 1s to Co 4p states. To simulate the absorption spectrum, calculated Co 4p DOS is convoluted with a Lorentzian having FWHM of 2.8 eV to account for the lifetime of Co 1s core hole generated during the absorption process. The result thus obtained is again convoluted with a Gaussian having FWHM of 2 eV to account for instrumental broadening and other contributions. This convoluted DOS generates almost all the features observed in the experimental spectrum except the pre-edge structures. This indicates that the pre-edge structures have origin beyond dipole approximation. Many workers [9,10,18] have attributed pre-edge structures observed in different transition metal oxides to 1s→3d quadrupole transitions. Müller et al. [19] have estimated the value of quadrupole contribution for K and L-edges. For K-edge it comes out to be less than 5% of dipole contribution for transition metal elements. Taking 2% of quadrupole contribution, by multiplying calculated Co 3d DOS by 0.02 and adding it to Co 4p DOS, and then convoluting the DOS with Lorenzian and Gaussian as above to take account of lifetime and instrumental broadening gives a good representation of both the pre-edge structures. The final convoluted DOS is shown by solid line in figure 1. This convoluted DOS clearly represents all the experimentally observed features including the pre-edge structures. We have observed small deviations in the positions of different structures between calculated and the experimental spectra. This may be due to the effects of energy dependence of transition matrix elements and instrumental broadening. Our results indicate that the shape of the main-edge is also fully generated. This indicates that the shape of the main-edge of $LaCoO_3$ is not much affected by the final state effect due to core hole potential and can be described under single particle approximation. The calculated $t_{2g}$ and $e_g$ DOS above the Fermi level, shown in the upper panel of inset of figure 1, shows that in the pre-edge region only $e_g^\uparrow$ and $e_g^\downarrow$ channels contribute and $t_{2g}$ band has negligible contribution. This indicates that first pre-edge feature has contribution mainly from $e_g^\uparrow$ channel and second one from $e_g^\downarrow$ channel. This result is different from that of Medarde et al. [9] where they have attributed two observed pre-edge structures to $t_{2g}$ and $e_g$ levels. The fact that calculated features are at slightly lower energies as compared to the experimental ones may be attributed to non-inclusion of orbital ordering in the DOS calculation as remarked earlier.

The post-edge region is mainly dominated by multiple scattering. To identify the nearest atoms that directly affect the Co 4p bands in higher energy levels we have calculated Co K-edge absorption spectra by cluster calculation using FEFF8.2 with different coordination spheres. These spectra together with the experimental one are shown in the upper panel of figure 2. It is seen from the figure that when multiple scattering (MS) calculations are carried out only over the first coordination shell of $CoO_6$ octahedron, a broad edge (solid line) is observed. The splitting of the spectrum with features E, F, and G takes place only when the next La atoms are included into the calculations (dashed line). Features F and G are not well resolved in the calculated spectra as seen in the experimental one. When further next neighbour Co atoms are added into the cluster these features become sharper. These calculations show that the details of post-edge structures are sensitive to the number of neighbouring atoms included in the calculations. The MS calculations do not give information about which particular electronic state of La atom is affecting most the Co 4p bands. To extract this information we have compared our calculated DOS with MS calculations. Although almost all the states included in the DOS calculations are hybridizing with Co 4p states in the post-edge region, it is the La 6p state that contributes most as shown in the lower panel of figure 2. Thus hybridization of Co 4p states with La 6p states changes the atomic character of 4p states and post-edge features seem to be representative of the strength of this hybridization. Therefore, MS calculations together with DOS calculations are useful in understanding how different states of the neighbouring atoms are affecting the Co 4p bands.

## La $L_1$-edge

$L_1$ absorption spectrum arises mainly due to transition of 2s electrons to unoccupied p-symmetric states. Due to highly localized nature of La 2s states the p-symmetric states of neighbouring atoms have negligible contribution to the absorption process. Thus it is the p-symmetric states of La atom that contribute to La $L_1$-edge spectrum, particularly La 6p states, as higher p-symmetric states lie at much higher energies. Open circles in the upper panel of figure 3 represent the experimental $L_1$-edge XA spectrum. This spectrum has mainly four structures denoted by A, B, C, and D.

The calculated DOS of La 6p states are shown by dotted lines in the lower panel of figure 3. The presence of finite DOS of O 2p at the Fermi level inspite of the system being insulating is a manifestation of non-inclusion of Jahn-Teller induced orbital ordering in the calculations as remarked earlier. It is evident from the figure that La 6p states are very much extended. To simulate the experimental XA spectrum we convoluted the calculated DOS by Lorentzian of FWHM ~ 4.7 eV to account for the core hole lifetime and a Gaussian of FWHM ~ 2 eV to account for the instrumental and other broadening effects. This convoluted DOS is shown by dashed line in the upper panel of figure 3. Although the features B, C, and D are seen, the structure A is not observed in the convoluted DOS. The position of La 5d band, shown by solid lines in the lower panel of figure 3, clearly indicates that the structure A can be generated by taking proper weightage of La 5d band; then it will correspond to a quadrupole transition. It is known that the contribution of quadrupole channel in absorption process increases with $Z^2$ where Z is the atomic number of the absorbing atom [19,20]. Thus it is expected that in this case with Z = 57 for La the amount of quadrupole contribution will be much more than that for Co K-edge with Z = 27. The final convoluted DOS obtained by taking 10% contribution of La 6d DOS in addition to La 6p DOS and broadening the whole with Lorentzian and Gaussian as above, is shown by the solid line in the figure. It is evident that all the features of the experimental spectrum including the structure A are well represented by this convoluted DOS. Although a small deviation in the position is observed on the lower energy side, the shape of the edge is generated reasonably well on including the quadrupole contribution. This deviation on the low energy side may be due to the stronger dependence of the quadrupole transition upon the energy dependent matrix element as compared to that for the corresponding dipole transition. This is expected due to strong angle dependence of quadrupole Hamiltonian in comparison to the dipole one. This fact has also been confirmed by MS calculations discussed below. The post-edge structures C and D can be seen in the convoluted DOS. The position of feature C matches with the experimental spectrum but that of D is on the lower energy side in the convoluted DOS. This may also be due to the transition matrix elements contribution to the absorption process.

MS calculations of La $L_1$-edge are shown in figure 4. The dotted and solid lines, respectively represent the absorption spectra when dipole-only and dipole + quadrupole contributions are taken into account. It is seen from the figure that a small shoulder in the low energy side of main-edge in dipole MS calculation develops into structure A when quadrupole contribution is included. This confirms contribution of the quadrupole channel to the structure A. Energy dependence of quadrupole contribution is shown in the inset of figure 4. It is seen that the quadrupole contribution is highly energy dependent in conformity with earlier assertion. It has a maximum contribution of ~ 11% around 6273 eV as compared to the dipole channel. To identify the effect of neighbouring atoms on the post-edge structures we have done MS calculations by considering different coordination shells. Results from these calculations are also shown in figure 4. It is seen that all the structures can be generated by including the first neighbours- oxygen atoms. This is in contrast to the post-edge structures in Co K-edge spectrum where structure E, F, and G (figure 2) develop only after including second neighbours- La atoms. The results for Co K-edge revealed that Co 4p states are affected most by La 6p states. Conversely, one would also expect the effect of Co 4p states on La 6p states. However, the calculations show that the effect of Co 4p states is not strong enough to create new structures in the post-edge region when neighbouring Co atoms are included in the calculations. Signature of such behaviour should be present in the calculated DOS. The results of the calculated partial DOS indicate that in the post-edge region the magnitude of O 2p DOS is more than three times that of Co 4p DOS. Thus it is the O 2p state that affects most the atomic character of La 6p states. This may be the reason why inclusion of neighbouring oxygen atoms is sufficient to generate all the post-edge structures. Addition of further neighbouring atoms in the cluster does not yield any new features, though the existing structures are made sharper.

### La $L_2$ and $L_3$ edges

For $L_2$ and $L_3$ edges the initial states are $2p_{1/2}$ and $2p_{3/2}$, respectively. Therefore, under dipole approximation $L_2$- and $L_3$- absorption take place due to transition of electron from $2p \rightarrow s$ and d-symmetric states. It is known that the contribution of $2p \rightarrow s$ transition is very small in comparison to $2p \rightarrow d$ transition and thus the former may be ignored. The experimental La $L_2$-edge XA spectrum, indicated by open circles, is shown in the upper panel of figure 5 and that of $L_3$-edge in the inset of the upper panel. These spectra show mainly three structures denoted by A, B, and C.

Calculated DOS shows that only Co 3d and La 5d have d-symmetric states above the Fermi level. Co 3d state does not contribute much to the XA process as it is localized on a different site. Thus only La 5d state would contribute to $L_2$ and $L_3$ XA processes. Calculated La 5d DOS is denoted by solid line in the lower panel of figure 5. To simulate the experimental XA spectra La 5d DOS is convoluted with Lorentzian and Gaussian having FWHM of 1 eV to account for the core hole lifetime and instrumental broadening [21], respectively. Convoluted La 3d DOS is indicated by dotted line in the upper panel of figure 5. It is clear that this DOS does not represent fully the observed experimental spectrum as it is much broader and its main peak splits into two. As it has been mentioned earlier there is about 10% contribution of the quadrupole channel in $L_1$-edge absorption spectrum. Thus it is expected that quadrupole channel will also contribute in $L_2$ and $L_3$ edges. In this case only La 4f state contributes to the quadrupole channel. La 4f DOS is shown with dotted line in the lower panel of figure 5. As in the case of $L_1$-edge, we have added 10% of La 4f DOS to La 5d DOS and convoluted with Lorentzian and Gaussian. Resulting convoluted DOS is denoted by solid line in the upper panel of figure 5. This is a better representative of the spectrum as it has only one peak at ~ 5897.7 eV. This confirms contribution of quadrupole channel in $L_2$ and $L_3$ XA processes. Features B and C observed in the experimental spectra are also seen in the convoluted DOS at slightly different energy positions.

MS calculations for La $L_2$-edge with different cluster sizes are shown in figure 6. As in the case of La $L_1$-edge all the post-edge structures are generated when only neighbouring oxygen atoms are included. On further including neighbouring Co atoms no new structures appear and the earlier structures become sharper. They are further developed on increasing the number of atoms in the cluster. Since the consideration of neighbouring oxygen atoms is sufficient to generate post-edge features, only O states will have major contribution in changing the atomic character of La 5 d states. This is in conformity with calculated DOS shown in the lower panel of figure 5. It thus follows that O 2p state has major contribution to the hybridization with La 5d and 4f states.

**Summary and concluding remarks**

The room temperature X-ray absorption studies at Co K- and La L-edges have been carried out on $LaCoO_3$. All the experimental spectra have been analysed using LDA+U density of states (DOS) and multiple scattering (MS) calculations. All the experimentally observed near edge features in Co K, La $L_1$, and $L_2$-$L_3$ edges spectra have also been present in the convoluted Co 4p, La 6p and La 5d densities of states, respectively. This indicates that the ground state obtained within single particle approximation is able to reproduce all major experimental features of the X-ray absorption spectra even in a highly correlated system like $LaCoO_3$. Presence of quadrupole contributions in the pre-edge region of Co K-edge and main-edge region of La L-edge has been seen. The two Co K-edge pre-edge structures have been attributed to Co 1s$\rightarrow e_g^\uparrow$ and $e_g^\downarrow$ quadrupole transitions. Such attribution will be helpful in understanding the temperature induced spin state transition depending on the occupancies of $e_g^\uparrow$ and $e_g^\downarrow$ channels. This work reports for the first time use of both MS and DOS calculations to understand the fine features of the spectra. It is seen to yield additional information from the experimental data. Thus we find that the hybridization of Co 4p state with La 6p state is crucial for understanding the post-edge structures in Co K-edge spectrum, whereas hybridization of La 6p and 5d states with O 2p state is important to generate the post-edge structures observed in La $L_1$ and $L_2$-$L_3$ edges spectra, respectively.

**Acknowledgements**

We would like to acknowledge N. P. Lalla and S. Bhardwaj for their help in XRD measurements. SKP thanks UGC-DAE CSR for financial support. AK thanks CSIR for senior research associate position (pool scheme) of Government of India.

**References**

*E-mail: sk_iuc@rediffmail.com

§E-mail: avp@csr.ernet.in

[1] P. M. Raccah and J. B. Goodenough, Phys. Rev. **155**, 932 (1967); V. G. Bhide, D. S. Rajoria, and Y. S. Reddy, Phys. Rev. Lett. **28**, 1133 (1972); V. G. Sathe, A. V. Pimpale, V. Siruguri, and S. K. Paranjpe, J. Phys.: Condens. Matter **8**, 3889 (1996); P. G. Radaelli and S.-W. Cheong, Phys. Rev. B **66**, 094408 (2002); V. P. Plakhty, P. J. Brown, B. Grenier, S. V. Shiryaev, S. N. Barilo, S. V. Gavrilov and E. Ressouche, J. Phys.: Condens. Matter **18**, 3517 (2006)

[2] S. Yamaguchi, Y. Okimoto, H. Taniguchi, and Y. Tokura, Phys. Rev. B **53**, R2926 (1996); J.-Q. Yan, J.-S. Zhou, and J. B. Goodenough, Phys. Rev. B **69**, 134409 (2004)

[3] J. B. Goodenough, J. Phys. Chem. Solids **6**, 287 (1957)

[4] R. H. Potze, G. A. Sawatzky, and M. Abbate, Phys. Rev. B **51**, 11501 (1995)

[5] G. Thornton, I. W. Owen, and G. P. Diakun, Phys.: Condens. Matter **3**, 417 (1991)

[6] M. Abbate, J. C. Fuggle, A. Fujimori, L. H. Tjeng, C. T. Chen, R. Potze, G. A. Sawatzky, H. Eisaki, and S. Uchida, Phys. Rev. B **47**, 16124 (1993)

[7] T. Saitoh, T. Mizokawa, A. Fujimori, M. Abbate, Y. Takeda, and M. Takano, Phys. Rev. B **55**, 4257 (1997)

[8] Z. Y. Wu, M. Benfatto, M. Pedio, R. Cimino, S. Mobilio, S. R. Barman, K. Maiti, and D. D. Sarma, Phys. Rev. B **56**, 2228 (1997)

[9] M. Medarde, C. Dallera, M. Grioni, J. Voigt, A. Podlesnyak, E. Pomjakushina, K. Conder, Th. Neisius, O. Tjernberg, and S. N. Barilo, Phys. Rev. B **73**, 054424 (2006)

[10] F. M. F. de Groot, M. Grioni, J. C. Fuggle, J. Ghijsen, G. A. Sawatzky, and H. Petersen, Phys. Rev. B **40**, 5715 (1989)

[11] J. M. Lawrence, M. L. den Boer, R. D. Parks, and J. L. Smith, Phys. Rev. B **29**, 568 (1984)

[12] M. A. Korotin, S. Yu. Ezhov, I. V. Solovyev, V. I. Anisimov, D. I. Khomskii, and G. A. Sawatzky, Phys. Rev. B **54**, 5309 (1996); I. A. Nekrasov, S. V. Streltsov, M. A. Korotin, and V. I. Anisimov, Phys. Rev. B **68**, 235113 (2003); K. Knízek, Z. Jirák, J. Hejtmánek, and P. Novák, J. Phys.: Condens. Matter **18**, 3285 (2006)

[13] S. K. Pandey, R. Bindu, P. Bhatt, S. M. Chaudhari, and A. V. Pimpale, Physica B, **365**, 45 (2005)

[14] S. Y. Savrasov, Phys. Rev. B **54**, 16470 (1996)

[15] J. P. Perdew, K. Burke, and M. Ernzerhof, Phys. Rev. Lett. **77**, 3865 (1996)

[16] S. K. Pandey, Ashwani Kumar, S. M. Chaudhari and A. V. Pimpale, J. Phys.: Condens. Matter **18**, 1313, (2006)

[17] A. L. Ankudinov, B. Ravel, J. J. Rehr, and S. D. Conradson, Phys. Rev. B **58**, 7565 (1998)

[18] J. E. Hans, R. A. Scott, K. O. Hodgson, S. Doniach, S. R. Desjardins, and E. I. Solomon, Chem. Phys. Lett. **88**, 595 (1982); A. Y. Ignatov, N. Ali, and S. Khalid, Phys. Rev. B **64**, 014413 (2001)

**Figure captions**

Figure 1  Experimental Co K-edge X-ray absorption spectrum (open circles) and convoluted Co 4p + 2% of Co 3d density of states (solid line). In the upper panel of the inset $t_{2g}$ (dotted lines) and $e_g$ (solid line) unoccupied density of states above the Fermi level are shown and in lower panel two pre-edge structures A and B observed in the experimental spectrum (open circles) and calculated spectrum (solid line) are shown.

Figure 2  In the upper panel Co K-edge experimental (open circles) and calculated X-ray absorption spectra using multiple scattering calculations at different coordination shells are shown. Solid, dashed, dotted, and dash dotted lines represent the calculated absorption spectra when Co-O, Co-O-La, Co-O-La-Co, and Co-O-La-Co-O atoms, respectively, are included in the calculations. In the lower panel Co 4p (solid lines) and La 6p (dotted lines) density of states per formula unit are shown.

Figure 3  In the upper panel experimental La $L_1$-edge X-ray absorption spectrum (open circles) and convoluted La 6p (dashed line) and La 6p + 10% of La 5d (solid line) density of states are shown. In the lower panel La 6p, 5d and O 2p density of states per formula unit are denoted by dotted, solid, and dashed lines, respectively.

Figure 4  La $L_1$-edge experimental (open circles) and calculated X-ray absorption spectra using multiple scattering calculations at different coordination shells are shown. Dotted and solid lines represent the calculated spectra when dipole and dipole+quadrupole contributions, respectively, are included in the calculations. Dashed, dash dotted, dash dot dotted, and short dashed lines represent the calculated absorption spectra when La-O, La-O-Co, La-O-Co-La, and La-O-Co-La-O atoms, respectively, are included in the calculations. In the inset quadrupole contribution to the absorption spectrum is shown.

Figure 5  In the upper panel experimental La $L_2$- ($L_3$- in the inset) edge X-ray absorption spectrum (open circles) and convoluted La 5d (dotted line) and La 5d + 10% of La 4f (solid line) density of states are shown. In the lower panel La 5d, 4f and O 2p density of states per formula unit are denoted by solid, dotted, and dashed lines, respectively.

Figure 6  La $L_2$-edge experimental (open circles) and calculated (solid line) X-ray absorption spectra using multiple scattering calculations. Dashed, dotted, dash dotted, and dash dot dotted lines represent the calculated absorption spectra when La-O, La-O-Co, La-O-Co-La, and La-O-Co-La-O atoms, respectively, are included in the calculations.

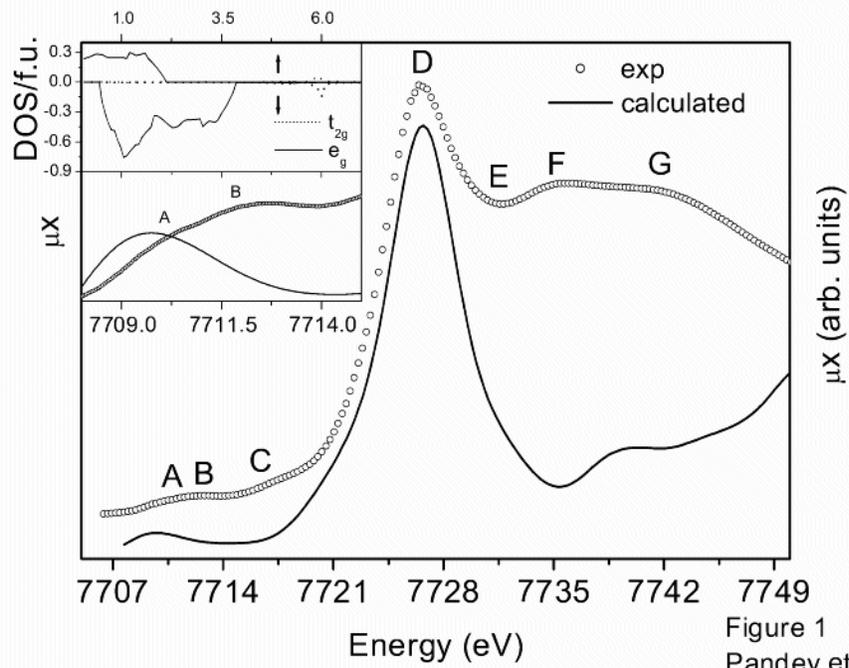

Figure 1
Pandey et al.

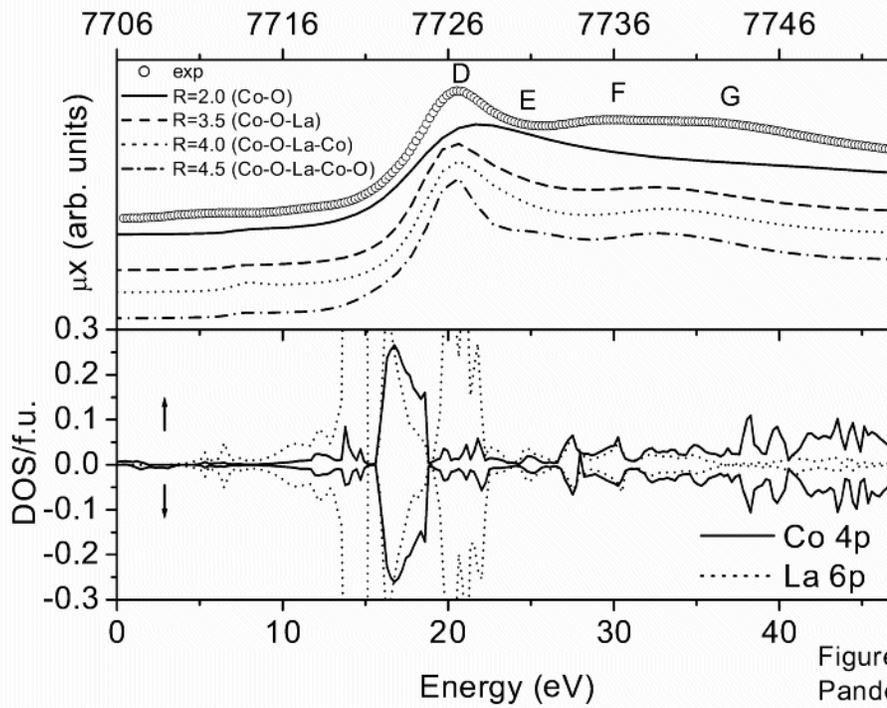

Figure 2
Pandey et al.

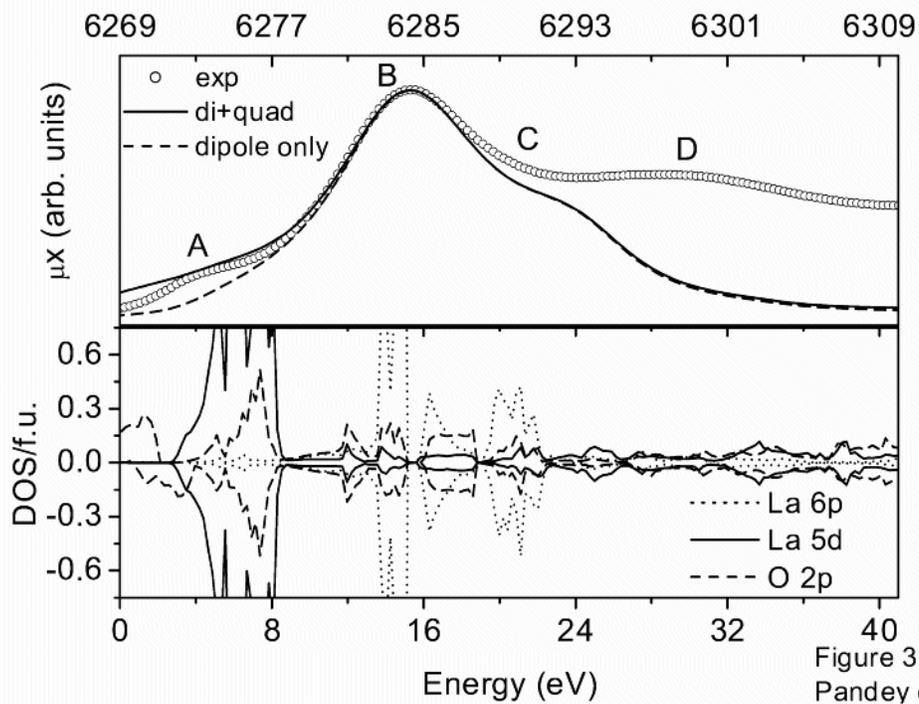

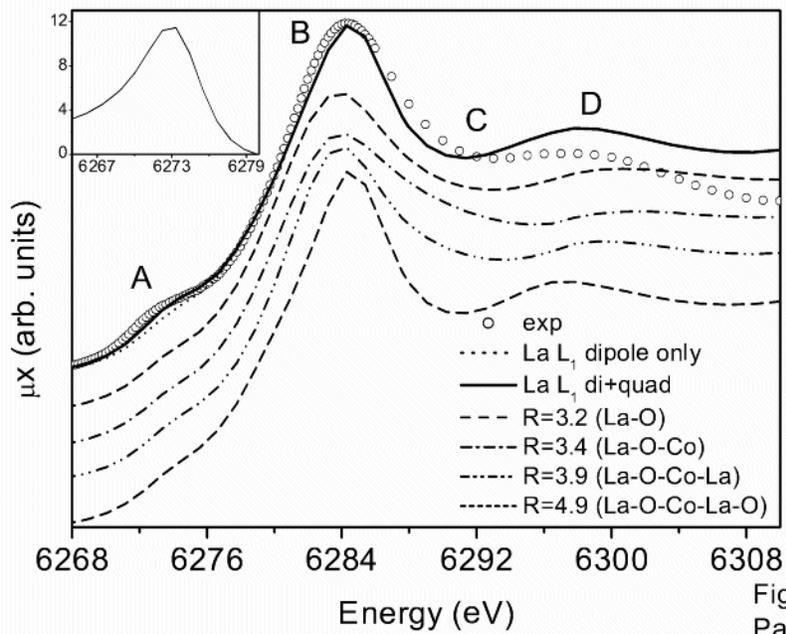

Figure 3 Pandey et al.

Figure 4 Pandey et al.

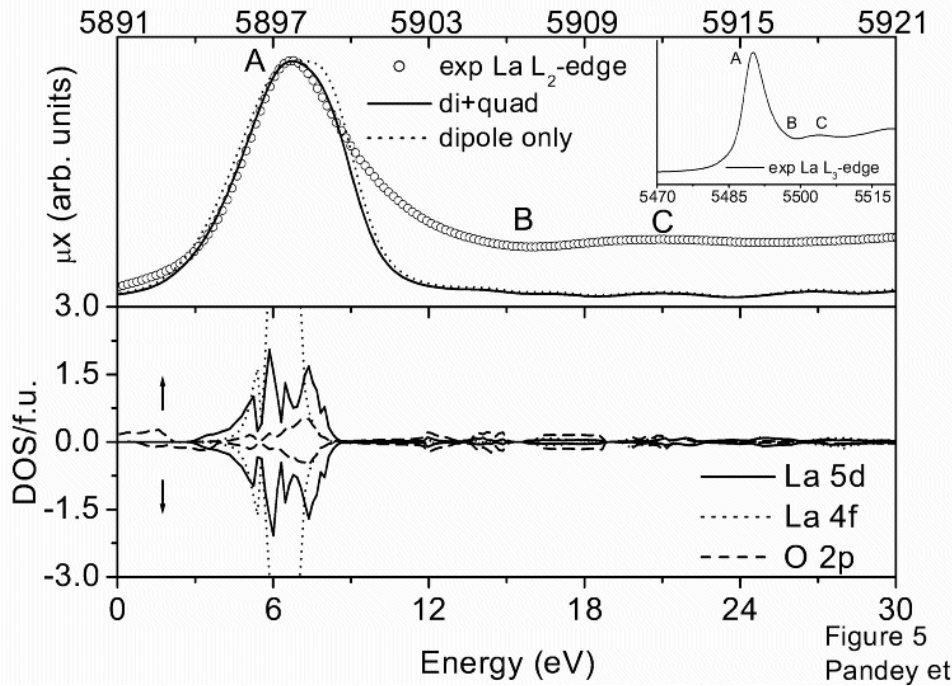

Figure 5
Pandey et al.

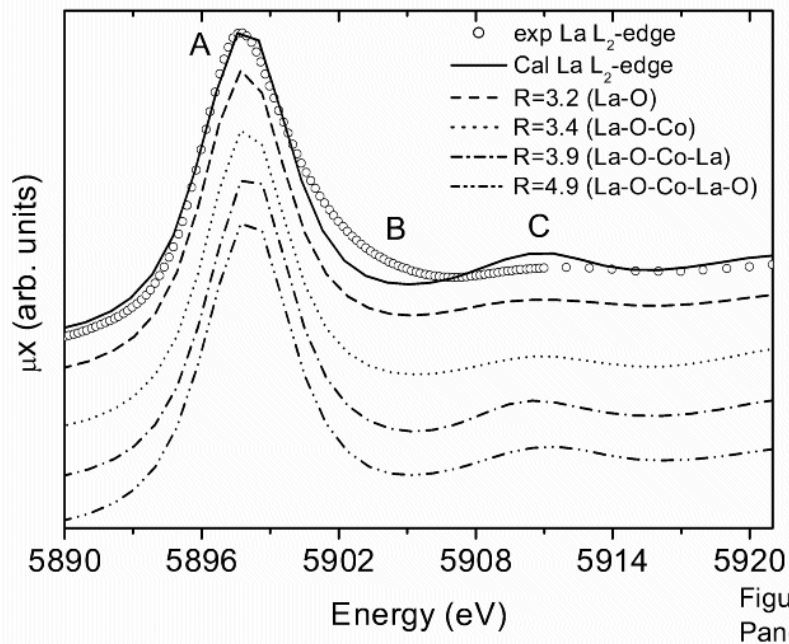

Figure 6
Pandey et al.

13